\documentclass[preprint,floatfix] {revtex4}
\newcommand{\rvec}{\mathrm {\mathbf {r}}} 
\newcommand{\rpvec}{\mathrm {\mathbf {r'}}} 
\usepackage{graphicx}
\usepackage{subfigure}
\begin{document}

\title{Density functional calculation of many-electron systems in Cartesian coordinate grid}
\author{Amlan K.\ Roy}
\email{akroy@iiserkol.ac.in, akroy@chem.ucla.edu} 
\affiliation{Division of Chemical Sciences, Indian Institute of Science Education and \\
Research (IISER), Block FC, Sector III, Salt Lake City, Kolkata-700106, India}

\begin{abstract}
A recently developed density functional method, within Hohenberg-Kohn-Sham framework, is used for 
faithful description of atoms, molecules in Cartesian coordinate grid, by using an LCAO-MO 
ansatz. Classical Coulomb potential is obtained by means of a Fourier convolution technique. All two-body 
potentials (including exchange-correlation (XC)) are constructed directly on real grid, while their 
corresponding matrix elements are computed from numerical integration. Detailed systematic investigation 
is made for a representative set of atoms/molecules through a number of properties like total energies, 
component energies, ionization energies, orbital energies, etc. Two nonlocal XC functionals (FT97 and PBE) are 
considered for pseudopotential 
calculation of 35 species while preliminary all-electron results are reported for 6 atoms using the LDA  
XC density functional. Comparison with literature results, wherever possible, exhibits near-complete 
agreement. This offers a 
simple efficient route towards accurate reliable calculation of many-electron systems in the Cartesian grid. 
Future prospect of this method is also discussed.
\end{abstract}
\maketitle

\section{Introduction}
Within density functional theory (DFT), ground-state energy of a many-electron system is conveniently divided 
into specific components as follows:
\begin{equation}
E[\rho(\rvec)] = T_{ni}[\rho(\rvec)]+V_{ne}[\rho(\rvec)] + V_{ee}[\rho(\rvec)] +\Delta T[\rho(\rvec)]
+\Delta V_{ee}[\rho(\rvec)]
\end{equation}
Here, from left to right, the terms in right-hand side denote kinetic energy of the non-interacting
electrons, nuclear-electron attraction, classical electron-electron repulsion, \emph{correction} to 
kinetic energy arising out of the interacting nature of electrons, and \emph{all} non-classical corrections
to the electron repulsion energy (including exchange and correlation), respectively. Now, using an orbital
expression for density, above equation may be recast as,
\begin{equation}
E[\rho(\rvec)]= \sum_i^N \left ( \left \langle \psi_i \left | \frac{1}{2}\nabla_i^2 \right | 
\psi_i \right \rangle - \left \langle \psi_i \left | \sum_k^M \frac{Z_k}{|\rvec_i-\rvec_k|} 
\right | \psi_i \right \rangle 
+ \left \langle \psi_i \left | \frac{1}{2} \int \frac{\rho(\rpvec)}{|\rvec_i -\rpvec|} 
\ d\rpvec \ \right | \psi_i  \right \rangle \right ) +E_{xc}[\rho(\rvec)]
\end{equation}
where N, M denote number of electrons and nuclei respectively, whereas density of a Slater determinantal 
wave function (exact for a non-interacting system) is simply $\rho=\sum_i^N \langle \psi_i|\psi_i \rangle$.

As in the Hartree-Fock (HF) case, Kohn-Sham (KS) DFT equations also cannot be solved exactly and recourse must
be taken to approximations. However, the required iterative process of solving KS equations is conceptually
very similar to that encountered for solving HF equations. Two major routes have been explored for practical
solution of a molecular KS equation. The \emph{real-space} method \cite{white89,
chelikowsky00,wang00,kronik06} involves an iterative mechanism for a
discretized KS equation on a real mesh using either of finite-difference, finite-element or wavelet 
technique. Typically this \emph{whole} molecular grid belongs to either uniform or refined uniform grids.
Some important advantages of this method are that (a) grid-based matrix representation offers highly structured banded
matrices (b) potential operator is diagonal in coordinate space whereas Laplacian operator is nearly local
(c) they are easily amenable to the so-called linear scaling methods. Usually these schemes require exceedingly
large grid points to capture the complete physical system of interest, which is essential to deliver physically
and chemically acceptable results. However with the introduction of higher order and multigrid techniques,
grid points could be considerably cut down without sacrificing much accuracy. 

The other alternative, which is more often used these days, relies on an expansion of eigenfunctions in terms 
of some suitable \emph{basis functions} such as Slater or Gaussian type functions (GTF), plane waves, numerical 
functions, augmented plane waves, linear muffin-tin orbitals, etc. Of these, GTFs are the most favorites, for 
they provide easy analytic routes towards relevant multi-center integrals, 
\begin{equation}
\psi_i (\rvec)= \sum_{\mu=1}^K C_{\mu i} \chi_{\mu}(\rvec), \ \ \ \ \ i=1,2,\cdots,K
\end{equation} 
A central problem common to almost all DFT approaches is that of obtaining the classical Coulomb ($V_c$) and XC
($V_{xc}$) potentials from electron density. In general, these cannot be obtained in any analytic form, 
and hence numerical treatments are necessary for evaluation of subsequent matrix elements. This is also true
for energy integrals associated with XC energy; $E_{xc}[\rho(\rvec)]= \int \epsilon_{xc} [\rho(\rvec)] \ 
d\rvec$. In order to circumvent this problem, in some molecular DFT implementations (see, 
\cite{amant90,andzelm92}, for example), 
electron density and XC potentials are \emph{also} expanded in some auxiliary Gaussian bases \emph{viz.,} 
$\rho(\rvec) \simeq \tilde{\rho}(\rvec)=\sum_i^{N_{\rho}} C_i^{\rho} f_i^{\rho} (\rvec),$ $V_{xc}(\rvec) \simeq 
\tilde{V}_{xc}(\rvec) = \sum_j^{N_{xc}} C_j^{xc} f_j^{xc} (\rvec)$, following some fitting procedure (so called discrete 
variational method) \cite{sambe75,dunlap79}, in addition to an MO expansion. This
facilitates an $N^3$ scaling.

A vast majority of modern DFT implementations employ the so-called atom-centered grid (ACG), pioneered by Becke 
\cite{becke88}, where a molecular grid is conveniently described in terms of some suitable 3D quadratures. The basic idea 
is to decompose real molecular space into overlapping atomic regions which are described by fuzzy Voronoi 
polyhedra. These discrete mono-centric atomic integrals can be individually approximated using some standard 
numerical techniques. Finally, summing these contributions with appropriate weight functions leads to the desired
molecular integration result. The atomic grid constitutes of a tensor product between radial part, defined
in terms of some quadrature formulas such as Gauss-Chebyshev, Gaussian, Euler-McLaurin, multi-exponential 
numerical, etc., \cite{gill93,murray93,treutler95,mura96,lindh01,gill03,chien06} and Lebedev angular 
quadratures (order as high as 131 has been reported, although usually much lower orders suffice; 59th order 
is the one most frequently used) \cite{lebedev92,lebedev94,lebedev1999}. Many variants of
original Becke integration scheme have been proposed thereafter, mainly to prune away some extraneous grid points
which could be quite beneficial.
Attempts were also made to perform such integrations by dividing whole space and invoking product Gauss rule to 
complete the subsequent integrations \cite{boerrigter88}. In a variational integration scheme, on the other hand, 
molecular space 
has been categorized into three different regions such as atomic spheres, excluded cubic region and interstitial 
parallelepiped \cite{pederson90}. In the Fourier transform Coulomb and multiresolution technique, Cartesian 
coordinate grid (CCG) was used in addition to ACG \cite{brown06,kong06}; former divides Gaussian shell pairs into 
``smooth" and ``sharp" categories on the basis of exponents while latter connects ACG and CCG by means of a 
divided-difference polynomial interpolation to translate density and gradients from latter to former.

Recently, DFT calculations \cite{roy08,roy08a,roy08b} of atoms and molecules have been performed using linear 
combination of Gaussian-type-orbital-expansion for KS molecular orbitals within CCG only. While no auxiliary basis 
set was invoked for charge density or XC potentials, localized atom-centered basis functions, MOs, electron 
density as well as classical two-electron Hartree and non-classical XC potentials were built on 3D real 
grid directly. The Coulomb potential was obtained from a Fourier convolution technique, involving a 
combination of FFT and inverse FFT, accurately and efficiently \cite{martyna99,minary02}. Analytical 
one-electron Hay-Wadt-type effective core potentials, made of a sum of GTFs, were used to 
represent the inert core electrons while energy-optimized truncated Gaussian bases were used for valence 
electrons \cite{wadt85,hay85a}. Detailed results were presented \cite{roy08,roy08a,roy08b} including total energy, 
energy components, 
orbital energy, potential energy curve, atomization energy for local density-approximated (LDA) 
\cite{vosko80} and non-local Becke exchange \cite{becke88a}+Lee-Yang-Parr correlation \cite{lee88}, to assess 
the performance and accuracy of this newly proposed method. Pseudopotential calculations on about 5 atoms and 53 
molecules for these above quantities produced practically identical results as those obtained from the 
widely used GAMESS quantum chemistry program \cite{schmidt93}, which, of course, works in ACG. It is well-known
that although LDA and BLYP functionals perform satisfactorily for many physical and chemical processes, in many
occasions they behave rather poorly and clearly there is genuine need for better functionals. In fact, construction of
accurate, proper XC functionals has been a very active, fertile area of research ever since the inception of KS 
formalism. The literature is vast and it is an ongoing process. Some of the other functionals in recent use are generalized
gradient expansion, hybrid functionals, meta functionals, orbital-dependent functionals, etc., (see, e.g., 
\cite{cramer04} for a lucid review). In this article, we employ two of them, \emph{viz.}, Filatov-Thiel (FT97) 
\cite{filatov97, filatov97a} and PBE \cite{perdew96} (for 8 atoms, 27 molecules), in order to extend the scope and 
applicability of this approach. These functionals have been used in many applications of many-electron systems with
reasonably good success. Also one of our main 
objectives to develop such a full CCG-based DFT procedure lies in the hope that LCAO-MO-DFT, which has enjoyed such a 
conspicuous success for enormous application for electronic structure of atoms/molecules in static case during past 
several decades, might also be equally successful for real-time dynamics studies (especially atoms/molecules in 
presence of an external field, such as a strong laser field, etc.) within the broad rubric of time-dependent (TD) DFT. 
Although considerable theoretical progress has been made for real-space dynamical studies of atoms/molecules under strong fields 
within TDDFT (see, for example, \cite{marques06}, for a review), several nontrivial problems are encountered for arbitrary 
atomic/molecular system. Extension in these cases is not straightforward, for they pose considerable challenge. On the other hand, 
extension 
of these above-mentioned LCAO-MO-based DFT approaches within ACG is quite difficult in the TD domain. In order to proceed further 
in that direction, here, for the first time, we report full all-electron calculation of several atoms using the basis
set method, completely in CCG. This constitutes an essential first step (the ``structure" part) much needed for real-time
TDDFT studies. To this end, total energies, component energies, orbital energies as well as HOMO energies from these full 
calculations are compared systematically with reference literature values. Section II gives a brief 
overview of the methodology used; discussion on our results are given in Section
III, while we end with a few concluding remarks in Section IV.

\section{Methodology}
The method employed in this work has been presented before \cite{roy08,roy08a,roy08b} in some detail. Hence will not be repeated here;
only essential details are given. Unless otherwise mentioned, atomic units implied henceforth.

In KS DFT, the problem of calculating total ground-state electronic energy of a many-electron system is 
transformed into solving the following single-particle KS equation,
\begin{equation}
\left[ -\frac{1}{2} \nabla^2 +v_{ext} (\rvec)+v_h(\rvec)+v_{xc}(\rvec) \right] \psi_i(\rvec) = 
\epsilon_i \psi_i(\rvec)
\end{equation}
Here $v_{ext}$ signifies the external potential in which electrons move, containing an electrostatic potential due 
to the presence of nuclei, but may also include other terms (in present occasion, none); $v_h$ denotes 
classical Coulomb potential arising because of the electron distribution; and finally $v_{xc}$ corresponds to a multiplicative
XC potential that depends on electron density, but not on wave functions explicitly.

As already mentioned, KS MOs are built from localized Gaussian type basis functions as in Eq.~(3). The LCAO-MO 
approach is quite efficient; for it can give very accurate results and also it provides basis for creation of new 
methods such as order-N, Green's function approaches, etc. Note, full self-consistent DFT procedure requires specification 
of basis functions; therefore a price is to be paid for efficiency in terms of loss of generality (in contrast to, for 
example, a plane wave method, where ``one basis fits all" philosophy works). Several important factors must be considered for 
building and 
choosing basis functions for a particular problem; two most important of them being (i) reduction of number of 
functions and (ii) ease of computation of relevant integrals. The electron density is described in terms of basis 
functions and corresponding one-body density matrix $P$ as,
\begin{equation}
\rho(\rvec) =\sum_{i=1}^{N} \sum_{\mu=1}^{K} \sum_{\nu=1}^{K} C_{\mu i} C_{\nu i} 
\chi_{\mu}(\rvec) \chi_{\nu}(\rvec)=
\sum_{\mu} \sum_{\nu} P_{\mu \nu}\ \chi_{\mu}(\rvec) \chi_{\nu}(\rvec)
\end{equation}
where $P_{\mu \nu}$ denotes an element of the density matrix. In a spin-unrestricted formalism, $\rho(\rvec)=
\rho^{\alpha} (\rvec)+ \rho^{\beta} (\rvec), P=P^{\alpha}+P^{\beta}$, and KS SCF wave function satisfies the
following set of equations, which are reminiscent of Pople-Nesbet equations in HF theory,
\begin{equation}
F^{\alpha} C^{\alpha} =S C^{\alpha} \epsilon^{\alpha},  \ \ \ \ \ \ \ \mathrm{and} \ \ \ \ \ \ \ 
F^{\beta} C^{\beta} =S C^{\beta} \epsilon^{\beta}
\end{equation}
with the orthonormality conditions,
\begin{equation}
(C^{\alpha})^{\dagger} S C^{\alpha} = I, \ \ \ \ \ \ \  \mathrm{and} \ \ \ \ \ \ \ 
(C^{\beta})^{\dagger} S C^{\beta} = I
\end{equation}
Here $C^{\alpha}$, $C^{\beta}$ are matrices containing MO coefficients, S is the atomic overlap matrix, and 
$\epsilon^{\alpha}, \epsilon^{\beta}$ are diagonal matrices of orbital eigenvalues.
$F^{\alpha}$, $F^{\beta}$ are KS matrices corresponding to $\alpha, \beta$ spins respectively, having matrix 
elements as,
\begin{equation}
F_{\mu \nu}^{\alpha}= \frac{\partial E_{KS}}{\partial P_{\mu \nu}^{\alpha}} =H_{\mu \nu}^{\mathrm{core}} 
+J_{\mu \nu}+ F_{\mu \nu}^{XC\alpha}, \ \ \ \ \mathrm{and} \ \ \ \ 
F_{\mu \nu}^{\beta}= \frac{\partial E_{KS}}{\partial P_{\mu \nu}^{\beta}} =H_{\mu \nu}^{\mathrm{core}} 
+J_{\mu \nu}+ F_{\mu \nu}^{XC\beta}
\end{equation}
Here $H_{\mu \nu}^{\mathrm{core}}$ represents the bare-nucleus Hamiltonian matrix that accounts for one-electron
energies, including contributions from kinetic energy plus nuclear-electron attraction. $J_{\mu \nu}$ term
refers to matrices from classical Coulomb repulsion whereas the third term arises from non-classical XC effects.
Last one remains the most difficult and challenging part of the whole SCF process. 

All one-electron integrals including overlap, kinetic-energy, nuclear-electron attraction as well as pseudopotential 
matrix elements are identical to those found in HF theory in Cartesian Gaussian functions and are generated by 
standard recursion algorithms. Here we employ angular-momentum dependent 
pseudopotentials as those from \cite{wadt85,hay85a}. Classical Hartree potential is computed by means of a Fourier 
convolution technique \cite{martyna99,minary02}, shown to be quite accurate and efficient for molecular modeling. 
This relies on a Ewald summation type decomposition of $1/r$ in terms of a pair of short-range (in terms of 
complimentary error function) and long-range (in terms of error function) contributions; former can be obtained 
analytically whereas the latter is computed directly from FFT of real-space values. 
\begin{eqnarray}
\rho(\mathbf{k}_g) & = & \mathrm{FFT}\{\rho(\mathbf{r}_g)\}  \\
v_H(\mathbf{r}_g) & = & \mathrm{FFT}^{-1}\{v_H^c(\mathbf{k}_g)\ \rho(\mathbf{k}_g) \}  \nonumber
\end{eqnarray}
Here $\rho(\mathbf{k}_g)$, the Fourier integral of density, is easily calculated from standard FFT, while 
$v_H^c(\mathbf{k}_g)$ signifies that of Coulomb interaction kernel in the grid which requires caution.
The nonlocal XC functionals of \cite{filatov97,filatov97a,perdew96} are used in this work; while for LDA calculations
homogeneous electron-gas correlation \cite{vosko80} is used. The gradient-dependent functionals are handled by 
using a finite-orbital expansion method \cite{pople92}, which allows one to bypass the calculation of difficult
density Hessians. In the end, XC matrix elements are evaluated as,
\begin{equation}
F_{\mu \nu}^{XC \alpha}= \int \left[ \frac{\partial f}{\partial \rho_{\alpha}} \chi_{\mu} \chi_{\nu} +
 \left( 2 \frac{\partial f}{\partial \gamma_{\alpha \alpha}} \nabla \rho_{\alpha} + 
    \frac{\partial f} {\partial \gamma_{\alpha \beta}} \nabla \rho_{\beta} \right) 
    \cdot \nabla (\chi_{\mu} \chi_{\nu}) \right] d\rvec
\end{equation}
where $\gamma_{\alpha \alpha} = |\nabla \rho_{\alpha}|^2$, 
$\gamma_{\alpha \beta} = \nabla \rho_{\alpha} \cdot \nabla \rho_{\beta}$, $\gamma_{\beta \beta} = 
|\nabla \rho_{\beta}|^2$. The advantage is that $f$ is a function \emph{only} of local quantities $\rho_{\alpha}$, 
$\rho_{\beta}$ and their gradients. Non-local functionals are implemented using the Density Functional Repository 
program \cite{repository}. There is no direct analytic route to evaluate two-electron matrix elements. 
Present work uses numerical integration for these in a 3D CCG covering a cubic box. Resulting matrix-eigenvalue
problem is solved accurately and efficiently by means of standard LAPACK library package \cite{anderson99}. 
Self-consistent solutions are obtained by imposing a tolerance of $10^{-6}$ for energy and eigenvalues whereas
$10^{-5}$ for potential.

\section{Results and discussion}
At first, we show FT97 and PBE results for 8 atoms and 27 molecules in Table I within the pseudopotential framework. 
Throughout the whole article, 
molecular geometries are used from those in NIST database \cite{johnson06}. We report non-relativistic ground-state
total, kinetic and potential energies for all these species using the effective core potential of \cite{wadt85,hay85a}. 
For sake of completeness, our integrated electron density is also given which can sometimes work as a rough indicator
of accuracy and quality attained in a given calculation. Several grid parameters were tried to check convergence, as in 
previous papers \cite{roy08,roy08a}. However as expected and observed, they produced very similar results in the present
occasion as well; discrepancies 
were rather very small from one set to other. In the end, $N_r=128, h_r=0.3$ seemed to be a very good reasonable choice, in 
keeping with our observations in \cite{roy08a}. In contrast to our earlier works, the GAMESS theoretical results 
could not be reported in this case, as results from these functionals are not available there. Excepting the lone case of Na, 
in all other 34 cases, FT97 total energies are found to be consistently lower than PBE values. Keeping in mind the performance 
of our results
for LDA and BLYP results in previous occasions, one can safely conclude that our current results are also 
equally accurate and trustworthy. 

\begingroup
\squeezetable
\begin{table}
\caption {\label{tab:table1}Kinetic $\langle T \rangle$, potential $\langle V \rangle$, total 
($\langle E \rangle$) energies and $N$ for several atoms and molecules. PBE and FT97 results are given in a.u. 
See text for details.} 
\begin{ruledtabular}
\begin{tabular}{lrrrrrrrr}
System     & \multicolumn{2}{c}{$\langle T \rangle$} & \multicolumn{2}{c}{$-\langle V \rangle$} & 
\multicolumn{2}{c}{$-\langle E \rangle$}   & \multicolumn{2}{c}{$N$}  \\
\cline{2-3}  \cline{4-5} \cline{6-7} \cline{8-9} 
              &  PBE      &  FT97     & PBE       & FT97      & PBE       &  FT97     & PBE      &  FT97      \\ 
\hline 
Na            &  0.06795  &  0.06966  &  0.25344  &  0.25482  &  0.18548  &  0.18516  &  0.99999 &  0.99999 \\
Mg            &  0.24078  &  0.24422  &  1.05699  &  1.07027  &  0.81621  &  0.82605  &  1.99999 &  1.99999 \\
Na$_2$        &  0.13774  &  0.14152  &  0.52488  &  0.53711  &  0.38714  &  0.39559  &  1.99999 &  1.99999 \\ 
NaH           &  0.52183  &  0.52627  &  1.26204  &  1.27956  &  0.74020  &  0.75329  &  1.99999 &  1.99999 \\
Si            &  1.33042  &  1.37824  &  5.09142  &  5.15100  &  3.76099  &  3.77276  &  3.99999 &  3.99999 \\
Mg$_2$        &  0.48713  &  0.49602  &  2.12361  &  2.14993  &  1.63648  &  1.65391  &  3.99999 &  3.99999 \\
AlH           &  1.16864  &  1.19789  &  3.69496  &  3.74500  &  2.52631  &  2.54710  &  3.99999 &  3.99999 \\
MgH$_2$       &  1.24003  &  1.27593  &  3.20349  &  3.26711  &  1.96346  &  1.99118  &  3.99999 &  3.99999 \\
P             &  2.32561  &  2.39816  &  8.74650  &  8.83346  &  6.42089  &  6.43530  &  4.99999 &  4.99999 \\
As            &  2.04560  &  2.10981  &  8.11083  &  8.18445  &  6.06524  &  6.07465  &  4.99999 &  4.99999 \\
SiH           &  1.87657  &  1.92371  &  6.23753  &  6.30523  &  4.36096  &  4.38152  &  4.99999 &  4.99999 \\
AlH$_2$       &  1.70807  &  1.75842  &  4.80730  &  4.87927  &  3.09923  &  3.12085  &  4.99999 &  4.99999 \\
S             &  3.63873  &  3.70741  & 13.66799  & 13.75933  & 10.02926  & 10.05193  &  6.00000 &  6.00000 \\
Al$_2$        &  1.32292  &  1.34896  &  5.20904  &  5.25881  &  3.88612  &  3.90985  &  5.99999 &  5.99999 \\
PH            &  2.88362  &  2.95298  &  9.90110  &  9.99256  &  7.01748  &  7.03959  &  5.99999 &  5.99999 \\
SiH$_2$       &  2.42339  &  2.47351  &  7.39314  &  7.47217  &  4.96975  &  4.99866  &  5.99999 &  5.99999 \\
Cl            &  5.50351  &  5.56491  & 20.38372  & 20.47315  & 14.88021  & 14.90824  &  7.00000 &  7.00000 \\
Br            &  4.17153  &  4.22828  & 17.29168  & 17.37615  & 13.12015  & 13.14787  &  6.99999 &  6.99999 \\
SH            &  4.22535  &  4.28816  & 14.86811  & 14.95915  & 10.64276  & 10.67100  &  7.00000 &  7.00000 \\ 
HSe           &  3.55021  &  3.60577  & 13.32383  & 13.40808  &  9.77362  &  9.80231  &  6.99999 &  6.99999 \\
PH$_2$        &  3.43226  &  3.49937  & 11.05733  & 11.15375  &  7.62507  &  7.65438  &  6.99999 &  6.99999 \\
SiH$_3$       &  3.03138  &  3.09386  &  8.54321  &  8.63382  &  5.51184  &  5.53996  &  6.99999 &  6.99999 \\  
HBr           &  4.73573  &  4.79470  & 18.47807  & 18.57128  & 13.74234  & 13.77657  &  8.00000 &  8.00000 \\
HI            &  3.59156  &  3.64696  & 15.55969  & 15.64856  & 11.96813  & 12.00160  &  7.99999 &  7.99999 \\
PH$_3$        &  3.99643  &  4.06590  & 12.24024  & 12.34634  &  8.24382  &  8.28044  &  7.99999 &  7.99999 \\
H$_2$S        &  4.83030  &  4.89575  & 16.09889  & 16.19906  & 11.26859  & 11.30331  &  7.99999 &  7.99999 \\
H$_2$Se       &  4.10439  &  4.16513  & 14.49460  & 14.59143  & 10.39021  & 10.42631  &  8.00000 &  8.00000 \\
SiH$_4$       &  3.53302  &  3.60707  &  9.75741  &  9.86882  &  6.22438  &  6.26175  &  7.99999 &  7.99999 \\ 
P$_2$         &  4.74887  &  4.80387  & 17.72258  & 17.81522  & 12.97371  & 13.01135  &  9.99999 &  9.99999 \\
S$_2$         &  7.51586  &  7.59269  & 27.64063  & 27.76097  & 20.12477  & 20.16828  & 12.00000 & 12.00000 \\
Se$_2$        &  6.14886  &  6.21184  & 24.55724  & 24.66671  & 18.40838  & 18.45487  & 11.99999 & 11.99999 \\
Br$_2$        &  8.47065  &  8.55429  & 34.76026  & 34.89602  & 26.28961  & 26.34173  & 13.99999 & 13.99999 \\
H$_2$S$_2$    &  8.65115  &  8.74098  & 29.99986  & 30.14139  & 21.34871  & 21.40041  & 13.99999 & 13.99999 \\
S$_3$         & 11.36056  & 11.45542  & 41.58894  & 41.74371  & 30.22838  & 30.28829  & 17.99999 & 17.99999 \\
P$_4$         &  9.95263  & 10.05211  & 35.89638  & 36.05954  & 25.94375  & 26.00742  & 19.99999 & 19.99999 \\
\end{tabular}                                                                               
\end{ruledtabular}
\end{table}
\endgroup

Now, Table II offers a comparison of our calculated ionization energies ($-\epsilon_{\mathrm{HOMO}}$) with literature 
results, for all the 27 molecules of Table I. In addition to the aforementioned FT97 and PBE results, here we have also appended
the BLYP results from \cite{roy08b}, for sake of comparison. Out of these, experimental values are not available for
6 species, and wherever possible they are adopted from \cite{afeefy05}. Three ionization energies, although quantitatively 
different from each other as expected, produce similar qualitative results. However all 3 values are rather quite low 
compared to the experimental data. As is well-known, a number of factors such as basis set, XC functional, relativistic
effects, etc., are responsible for this discrepancy. This does not, however, interfere with the main objective of this work
directly and may be taken up later in our future studies. It is worth mentioning here that, none of these 3 functionals
lead to ionization energies for these molecules as good (or even close to) as those from LBVWN (reported in \cite{roy08b}). 
For a moderate set of
atoms and molecules, this latter XC combination showed significant improvements in HOMO energies over LDA, BLYP results and now 
FT97, PBE functionals as well. 

\begingroup
\squeezetable
\begin{table}
\caption {\label{tab:table2}Comparison of $-\epsilon_{\mathrm{HOMO}}$ energies (in a.u.) for some 
molecules calculated using BLYP, FT97 and PBE XC functionals, with literature data. BLYP and experimental results 
are taken from refs. \cite{roy08b} and \cite{afeefy05} respectively. See text for details.}
\begin{ruledtabular}
\begin{tabular}{lrrrrlrrrr}
Molecule & \multicolumn{4}{c}{$-\epsilon_{\mathrm{HOMO}}$} & Molecule & \multicolumn{4}{c}{$-\epsilon_{\mathrm{HOMO}}$} \\
\cline{2-5}  \cline{7-10} 
           &   BLYP   &   FT97   &  PBE     &  Expt.   &         &    BLYP  &  FT97    &   PBE    &    Expt.  \\ 
\hline
Na$_2$     &  0.1002  &  0.0862  &  0.0952  &  0.1798  &  NaH    &  0.1421  &  0.1231  &  0.1324  &   ---     \\
Mg$_2$     &  0.1530  &  0.1293  &  0.1395  &  ---     &  AlH    &  0.1715  &  0.1362  &  0.1521  &   ---     \\
MgH$_2$    &  0.2221  &  0.1857  &  0.2024  &  ---     &  SiH    &  0.1597  &  0.1305  &  0.1514  &  0.2900   \\
AlH$_2$    &  0.1631  &  0.1253  &  0.1468  &  ---     & Al$_2$  &  0.1400  &  0.1161  &  0.1318  &  0.1984   \\
PH         &  0.2133  &  0.1786  &  0.2012  &  0.3730  & SiH$_2$ &  0.2027  &  0.1619  &  0.1808  &  0.3278   \\ 
SH         &  0.2174  &  0.1809  &  0.1994  &  0.3830  & HSe     &  0.2057  &  0.1746  &  0.1903  &  0.3618   \\
PH$_2$     &  0.2111  &  0.1741  &  0.1970  &  0.3610  & SiH$_3$ &  0.1969  &  0.1590  &  0.1825  &  0.2990   \\
HBr        &  0.2603  &  0.2249  &  0.2439  &  0.4292  & HI      &  0.2432  &  0.2122  &  0.2302  &  0.3817   \\
PH$_3$     &  0.2287  &  0.1854  &  0.2054  &  0.3627  & H$_2$S  &  0.2190  &  0.1815  &  0.2006  &  0.3843   \\
H$_2$Se    &  0.2075  &  0.1731  &  0.1916  &  0.3635  & SiH$_4$ &  0.3156  &  0.2702  &  0.2919  &  0.4042   \\
P$_2$      &  0.2526  &  0.2206  &  0.2360  &  0.3870  &  S$_2$  &  0.2023  &  0.1594  &  0.1781  &  0.3438   \\
Se$_2$     &  0.1951  &  0.1576  &  0.1749  &  0.3160  & Br$_2$  &  0.2451  &  0.2115  &  0.2285  &  0.3865   \\
H$_2$S$_2$ &  0.2288  &  0.1928  &  0.2103  &  0.3418  & S$_3$   &  0.2294  &  0.1985  &  0.2143  &  ---      \\
P$_4$      &  0.2575  &  0.2369  &  0.2525  &  0.3432  & ---     &   ---    &   ---    &   ---    &  ---      \\
\end{tabular}                                                                               
\end{ruledtabular}
\end{table}
\endgroup

\begingroup
\squeezetable
\begin{table}
\caption {\label{tab:table3} Energy components as well orbital energies for several atoms using Cartesian grid.
All-electron calculations with LDA XC functionals using STO-3G basis set are given along with those obtained from 
reference GAMESS program. See text for details.} 
\begin{ruledtabular}
\begin{tabular}{lrrrrrr}
Quantity  & This work  & Ref.~\cite{schmidt93} & This work & Ref.~\cite{schmidt93} & This work & Ref.~\cite{schmidt93}  \\ 
\cline{2-3}  \cline{4-5} \cline{6-7} 
             & \multicolumn{2}{c}{Li} & \multicolumn{2}{c}{Be} & \multicolumn{2}{c}{B}  \\
\hline 
$\langle T \rangle $       &    7.38213  &    7.38212  &    14.84418 &    14.84419  &    25.30018 &    25.30018 \\ 
$\langle V^{ne} \rangle $  & $-$17.11549 & $-$17.11549 & $-$34.07189 & $-$34.07189  & $-$58.14361 & $-$58.14361 \\
$\langle E_h \rangle $     &    4.24103  &             &     7.64382 &              &    12.66071 &             \\  
$\langle E_x \rangle $     & $-$1.57407  &             & $-$2.40486  &              & $-$3.47888  &             \\  
$\langle E_c \rangle $     & $-$0.15489  &             & $-$0.23242  &              & $-$0.30248  &             \\  
$\langle V^{ee} \rangle $  &    2.51207  &    2.51207  &    5.00654  &     5.00656  &    8.87935  &     8.87936 \\
$\langle V \rangle $       & $-$14.60343 & $-$14.60342 & $-$29.06535 & $-$29.06533  & $-$49.26426 & $-$49.26425 \\
$\langle E \rangle $       & $-$7.22130  & $-$7.22130  & $-$14.22116 & $-$14.22114  & $-$23.96408 & $-$23.96406 \\
$N$                        &    2.99999  &    2.99999  &     3.99999 &     3.99999  &     4.99999 &     4.99999 \\
$\epsilon^{\alpha}_{1s}$   & $-$1.7289   & $-$1.7289   & $-$3.5755   & $-$3.5754    & $-$6.0874   & $-$6.0873   \\
$\epsilon^{\alpha}_{2s}$   & $-$0.0815   & $-$0.0815   & $-$0.1288   & $-$0.1288    & $-$0.2053   & $-$0.2053   \\
$\epsilon^{\alpha}_{2p_x}$ &             &             &             &              & 0.0224      &  0.0224     \\
$\epsilon^{\beta}_{1s}$    & $-$1.7139   & $-$1.7139   & $-$3.5755   & $-$3.5754    & $-$6.0722   & $-$6.0721   \\
$\epsilon^{\beta}_{2s}$    &             &             & $-$0.1288   & $-$0.1288    & $-$0.1578   & $-$0.1578   \\
\hline 
             & \multicolumn{2}{c}{C} & \multicolumn{2}{c}{N} & \multicolumn{2}{c}{O}  \\
\cline{2-3}  \cline{4-5} \cline{6-7} 
$\langle T \rangle $       & 37.92456    & 37.92456    & 53.66407     &  53.66407    &  73.44497    &  73.44497    \\ 
$\langle V^{ne} \rangle $  & $-$88.64983 & $-$88.64983 & $-$127.32649 & $-$127.32649 & $-$176.32432 & $-$176.32432 \\
$\langle E_h \rangle $     &  18.78009   &             &  26.67740    &              &  37.46227    &              \\  
$\langle E_x \rangle $     & $-$4.64014  &             & $-$5.98724   &              & $-$7.49030   &              \\  
$\langle E_c \rangle $     & $-$0.36805  &             & $-$0.43478   &              & $-$0.54395   &              \\  
$\langle V^{ee} \rangle $  &  13.77190   &             &  20.25538    &  20.25536    &  29.42801    &  29.42799    \\
$\langle V \rangle $       & $-$74.87793 &             & $-$107.07111 & $-$107.07113 & $-$146.89631 & $-$146.89634 \\
$\langle E \rangle $       & $-$36.95337 & $-$36.95339 & $-$53.40704  & $-$53.40701  & $-$73.45134  & $-$73.45137  \\
$N$                        &    5.99999  &    5.99999  &     6.99999  &     6.99999  &     7.99999  &     7.99999  \\
$\epsilon^{\alpha}_{1s}$   & $-$9.4882   & $-$9.4879   & $-$13.6312   & $-$13.6311   & $-$18.3331   & $-$18.3330   \\
$\epsilon^{\alpha}_{2s}$   & $-$0.3970   & $-$0.3970   & $-$0.6152    & $-$0.6153    & $-$0.7538    & $-$0.7537    \\
$\epsilon^{\alpha}_{2p_x}$ & $-$0.0675   & $-$0.0676   & $-$0.1671    & $-$0.1672    & $-$0.1941    & $-$0.1942    \\
$\epsilon^{\alpha}_{2p_y}$ & $-$0.0648   & $-$0.0649   & $-$0.1671    & $-$0.1672    & $-$0.1941    & $-$0.1942    \\
$\epsilon^{\alpha}_{2p_z}$ &             &             & $-$0.1671    & $-$0.1672    & $-$0.1085    & $-$0.1084    \\
$\epsilon^{\beta}_{1s}$    & $-$9.4573   & $-$9.4572   & $-$13.5837   & $-$13.5836   & $-$18.2972   & $-$18.2971   \\
$\epsilon^{\beta}_{2s}$    & $-$0.2923   & $-$0.2924   & $-$0.4449    & $-$0.4450    & $-$0.6301    & $-$0.6302    \\
$\epsilon^{\beta}_{2p_x}$  &             &             &              &              & $-$0.0378    & $-$0.0379    \\
\end{tabular}                                                                               
\end{ruledtabular}
\end{table}
\endgroup

So far all the results presented using our method employed some sort of effective core potentials to incorporate the 
effects of frozen core electrons; no investigation has been made for the so-called ``full" all-electron calculations. 
Although pseudopotential studies are advantageous for larger systems, especially those containing one or more heavier atoms 
where full calculations could be expensive, latter are very desirable otherwise, for they can provide more detailed and 
also more accurate results. Thus they add valuable insights into a particular problem. As long as the 
cost accuracy ratio permits, these are the preferred choices for most chemical and physical studies. In an attempt to 
deal with such situations, as a very first step, in Table III, some representative preliminary all-electron results 
are given for a set of 6 atoms using this approach to assess its level of performance and effectiveness in the said domain.
For all these, we use LDA XC potential, STO-3G basis set and $N_r=128, h_r=0.3$. We are not aware of any other attempts
where such studies have been made within the LCAO framework, using CCG only. Besides the point mentioned above in this
paragraph, there are other important motivations for this case study, which have been elaborated in Section I. These are all
open-shell systems and a thorough comparison with the GAMESS program is made for all of them, using same basis set as well
as same XC functional. Following quantities are reported: kinetic energy $\langle T \rangle$, nuclear-electron attraction
energy $\langle V_{ne} \rangle$, classical Hartree energy $\langle E_h \rangle $, exchange energy $\langle E_x \rangle$, 
correlation energy $\langle E_c \rangle$, two-electron potential energy $\langle V_{ee} \rangle $, total potential energy 
$\langle V \rangle$, total electronic energy $\langle E \rangle$, total integrated electron density $N$ as well as all the
$\alpha$- and $\beta$-spin orbital energies. Note that literature results employ Euler-McLaurin and Gauss-Legendre 
quadratures for radial and angular integrations respectively. The default grid option is used for all these reference results.
Individual Coulomb repulsion and XC energies from literature could not be cited as GAMESS output does not report
those. Quite clearly, for all these quantities, agreement with literature results is excellent (very similar accuracy, as 
we observed in the pseudopotential case before); present results practically coincide with the reference values. It is 
well-known that STO-3G basis set employed here is not a very accurate one, and used here only for the demonstration purposes. 
Certainly better basis sets needs to be 
used for realistic calculations. These, as well as the molecular case, may be considered in future communications. However, 
the main motivation, at this stage, was to establish the validity and feasibility of this approach in the context of full 
electronic structure calculation of many-electron systems.

\section{Concluding remarks}
Many-electron systems have been studied by LCAO-MO-DFT in CCG. Both atoms and molecules (small as well as medium) were
considered. Results have been presented for pseudopotential and all-electron calculations. For the former we employed FT97 and PBE
XC functionals, as a follow-up of our previous work in this direction which further consolidates the success of this approach.
For the latter, exploratory preliminary results were given using some rudimentary basis set within the homogeneous-electron
gas approximation; this further extends the scope and applicability of the current scheme. The basis set, MOs, electron density 
and various
potentials were generated in a CCG encapsulating a cubic box. Hartree potential was conveniently computed via a Fourier convolution
method. Two-body matrix elements were obtained by direct numerical integration. Detailed comparison has been made with literature 
results, wherever possible, for a variety of quantities such as total energy, component energy, orbital energy, ionization 
energy, etc. Agreement 
has been extremely good; present results are almost indistinguishable from reference ACG-based DFT values (out of 6 atoms, 
largest absolute deviation in total energy from literature values is 0.00003 a.u. only). However, it would be necessary to incorporate 
better and more practical basis sets as well as XC functionals for more meaningful physical, chemical applications. In essence, 
this present study 
confirms the fact that electronic structure calculation of many-electron systems can be performed within LCAO-MO-based DFT, 
using Gaussian basis sets, very accurately and efficiently through CCG, offering virtually same accuracy as ACG. 

\section{acknowledgment}
Prof.~S.~I.~Chu at the University of Kansas, Lawrence, KS, USA, is thanked for computational resource facilities. I
greatly appreciate the support provided by IISER, Kolkata's colleagues and staff members.

\bibliographystyle{unsrt}
\bibliography{refn}
\end{document}